# Neural-Embedded Graphical Model for Self-Consistent Hierarchical Upscaling of Complex Composites

Nuo Xu[1], Shaohua Chen[2*]

[1] College of Mechanics and Engineering Science, Hohai University, Nanjing, 211100, China

[2] College of Energy and Power Engineering, Nanjing University of Aeronautics and Astronautics, Nanjing, 210016, China

**Abstract**

A persistent challenge in computational physical modeling is the substantial disparity between the characteristic length scales of microstructures and macroscopic structural components. Multiscale modeling has been widely adopted to bridge this gap by coupling methodologies tailored to different scales. However, conventional approaches, such as asymptotic homogenization (bottom-up) and submodeling (top-down), often entail rigorous mathematical prerequisites or intricate interfacing procedures. To address these limitations, we introduce a fully scalable neural-embedded graphical model (NEGM) that provides a unified framework for the progressive upscaling of highly heterogeneous composite materials. Specifically, NEGM encodes all microstructure- and material-related complexities into constituent neural network blocks, which are then organized into a hypergraph to simulate progressively larger domains. Extensive numerical benchmarks demonstrate that NEGM reliably predicts the physical responses of 2D and 3D composites exhibiting strong material nonlinearity, arbitrary boundary conditions, and irregular geometries. Crucially, because NEGM relies solely on neural network training and inference, it offers a scale-invariant formulation. This enables iterative application of NEGM to upscale from the microscale to arbitrarily large scales, circumventing the need for complex interfacing protocols between disparate modeling frameworks. We validate this progressive upscaling strategy on a large mosaic composite domain, showing that the accumulated error can be effectively contained provided the constituent blocks achieve sufficiently high prediction accuracy. Our findings suggest that artificial neural networks not only enhance the efficiency of direct single-scale simulations, as previously demonstrated, but also provide a clean and elegant pathway toward streamlined multiscale modeling.

[*] Corresponding author. E-mail: csh990729@163.com (Shaohua Chen)

**Keywords**: artificial intelligence, physical modeling, complex graph, composite material, finite element method

## 1. Introduction

The rapid advancement of scientific machine learning (SciML) has catalyzed a revolution in the physical modeling of materials[1,2]. Artificial intelligence (AI), and specifically artificial neural networks (ANNs), has become a heated topic in this field due to its extraordinary capacity to approximate the highly non-linear, high-dimensional physics governing complex engineering systems[3,4]. Over the past two decades, the intersection of machine learning and physical modeling has produced a rich body of research demonstrating that ANNs can effectively learn complex relationships of the physical variables within a physical system.

Composite materials have become indispensable in aerospace, automotive, civil, and energy applications due to their exceptional strength-to-weight ratios, tailorable mechanical properties, and design flexibility[5]. Traditional finite element method (FEM), while robust, often require prohibitive computational resources when modeling complex microstructures, progressive damage, or multiscale phenomena of composite material[6]. This computational bottleneck has motivated the development of ANN-based surrogate models that can dramatically accelerate simulation workflows of composite media while maintaining acceptable accuracy[3]. Specifically, ANNs are increasingly employed in composite materials for constitutive modeling, facilitating the discovery of unknown laws and the creation of efficient numerical relations that can replace expensive traditional simulations[7–9].

Early work has established the feasibility of using ANN to directly predict responsive fields of composite medium from constituent parameters and microstructures[10–13], more recent advances have leveraged deep learning architectures, e.g., convolutional neural networks (CNNs), recurrent neural networks (RNNs), and physics-informed neural networks (PINNs), to tackle increasingly sophisticated problems in composite material researches, such as full-field stress prediction[14,15], multiscale homogenization[16], and inverse design[17,18]. This AI-based physical modeling paradigm of composite material is essentially centered on using AI to solve partial differential equations (PDEs), an area known as AI4PDEs[19–21]. Modern frameworks such as Physics-Informed Neural Networks (PINNs) directly embed governing equations into the network's loss function, allowing the model to solve forward and inverse problems while respecting fundamental physical symmetries and conservation laws[22–24]. Furthermore, operator learning approaches, including the Fourier Neural Operator (FNO) and DeepONet, aim to learn mappings between infinite-dimensional function spaces, enabling the rapid prediction of entire solution families rather than individual instances[25,26].

Despite these developments, current AI-based PDE solvers often face difficulties in dealing with arbitrary boundary conditions, irregular geometries, and higher-dimensional domains[27,28]. Standard PINNs often treat each spatial point individually and are not intrinsically informed of the domain geometry, which frequently necessitates costly retraining for every new shape. Convolutional architectures like CNNs and FNOs often rely on structured grids, making them difficult to apply to the unstructured meshes and irregular boundaries typical of real-world 3D components. Moreover, high-dimensional problems (3D spatial plus time) still suffer from a curse of dimensionality that can render conventional physics-informed training computationally prohibitive[25].

Although a series of works have targeted these difficulties, such as the use of point-cloud based operators and variational formulations, they often focus on linear problems and remain limited to Dirichlet boundary conditions (BCs)[29–31]. For example, many frameworks prioritize the hard enforcement of Dirichlet BCs while treating Neumann or Robin conditions only as "soft" constraints in the loss function, which can lead to poor convergence and solution inconsistency[29,30]. Additionally, some cutting-edge homogenization frameworks that integrate neural operators are currently restricted to linear settings, leaving the high-fidelity modeling of complex non-linear responses a significant challenge[31,32].

Finally, current AI-based algorithms are often single-scale solving algorithms, and the challenge of building simple, concise multi-scale modeling methods for extremely large heterogeneous material domains remains largely unattended. While traditional methods like $FE^2$ offer accurate multiscale descriptions, their computational cost at the 3D level is extreme, requiring an RVE analysis for each integration point[33,34]. Others use ANN to replace the most time-consuming part of the conventional modeling method to strongly accelerate the computational speed[35,36]. Existing AI-driven multiscale approaches are often restricted to 2D domains or rely on purely data-driven paradigms that lack the robustness of physics-based structures. There is a critical need for multiscale frameworks that can efficiently couple local microstructural responses with larger scale behaviors, such as through domain decomposition, to manage the complexity of large-scale heterogeneous systems[37–39].

Although AI-based modeling is often criticized for being data-driven rather than physics-driven, this characteristic can also be viewed as an advantage. By decoupling the modeling process from conventional physical constraints, AI modeling is not explicitly constrained by traditional scale-bridging procedures and can therefore alleviate some of the computational challenges associated with high-dimensional multiscale simulations, including the curse of dimensionality and length-scale disparities[40,41]. As a result, AI provides a unified framework for linking composite microstructures to material behavior across multiple scales[42,43]. For example, if small-scale AI models can be used to generate training data for larger-scale models, a hierarchical upscaling chain can be established. Each model is trained using data produced at the preceding scale, resulting in a recursively constructed multiscale framework. Such an approach has the potential to form a self-consistent multiscale modeling paradigm that bridges microscopic and macroscopic length scales.

In this work, we propose a neural-embedded graphical model (NEGM) as a self-consistent multiscale modeling framework. NEGM integrates a number of building-block ANNs which are associated with a specific length scale, into a unified inference architecture. This design allows the model to flexibly handle arbitrary boundary conditions and complex geometries. By leveraging this structure, NEGM can generate response data at a given scale to train the next-scale ANNs, thereby forming a hierarchical data-generation and training loop. Repeating this process enables progressive upscaling across length scales, leading to a fully AI-driven multiscale modeling paradigm.

The remaining of the text is organized as follows. In section 2, we introduce the architecture and algorithmic details of NEGM; in section 3, we demonstrate the usage of NEGM for 2D and 3D simulation domain with complex microstructure, arbitrary BCs and exotic domain geometry; in section 4, we introduce the self-consistent multiscale paradigm based on NEGM to solve the large composite simulation problem; in section 5, we conclude this work with a few remarks.

## 2. The architecture of NEGM

For convenience of description, we take steady-state thermal conduction as example to lay the theoretical foundation of the NEGM method, although the framework of NEGM can be tailored for arbitrary physical system. The governing equation is $\nabla \cdot \{\boldsymbol{\kappa}[T(\boldsymbol{x})]\nabla T\} = f(\boldsymbol{x})$, where $T$ is the temperature field to be determined, $\boldsymbol{\kappa}[T(\boldsymbol{x})]$ is the nonlinear conductivity tensor. Let $\Omega \subset \mathbb{R}^d (d = 2{,}3)$ be a bounded Lipschitz domain representing a heterogeneous composite material, it is assumed to partitioned into $M$ non-overlapping subdomains,

$$\Omega = \bigcup_{m=1}^{M} \Omega_m, \quad \Omega_i \cap \Omega_j = \emptyset \text{ for } i \neq j$$

Each subdomain $\Omega_m$ is characterized by a conductivity tensor $\boldsymbol{\kappa}_m \in \mathbb{R}^{d\times d}$, and all of subdomains form a piecewise conductivity tensorial field $\boldsymbol{\kappa}(\boldsymbol{x}) = \boldsymbol{\kappa}_m$ when $\boldsymbol{x} \in \Omega_m$. For common scenarios, $\boldsymbol{\kappa}(\boldsymbol{x}) \in L^\infty(\Omega; \mathbb{R}^{d\times d})$, $\boldsymbol{\kappa}(\boldsymbol{x})$ is symmetric and uniform ellipticity is satisfied. These assumptions allow discontinuities of $\boldsymbol{\kappa}(\boldsymbol{x})$ across material interfaces.

To solve the above Poisson's equation via numerical approach such as FEM, it is necessary to convert the differential form into integral form to take advantage of high-performance linear algebra algorithms. Multiplying the Poisson's equation by a test function and integrating by parts yield $\int \boldsymbol{\kappa}[T(\boldsymbol{x})]\nabla T \cdot \nabla v \, d\boldsymbol{x} = \int \boldsymbol{f} v \, dx$, which can be discretized into matrix form,

$$\mathbf{KT} = \mathbf{F}$$

Here, $\mathbf{K}$ is the global stiffness matrix, $\mathbf{T}$ is the vectorial nodal value of the physical field and $\mathbf{F}$ is the heat flux vector. For heterogeneous and nonlinear material, the stiffness matrix may depend on internal microstructure and temperature field $T$. Under such circumstances, FEM usually employs iterative approach such as the Newton-Raphson method to determine temperature field.

Here, we propose a strategy of using ANN to surrogate the physical behavior of an element in place of complicated physical relation governed by PDEs. Resorting to the universal approximation capability of ANN[44–47], we can introduce a divergence neural network $\mathcal{F}^{\nabla}$ for divergence operator, a conductivity neural network $\mathcal{G}^{\boldsymbol{\kappa}(\{T\})}$ for the temperature dependent conductivity tensor, and a gradient neural network $\mathcal{H}^{\nabla\{T\}}$ for the gradient operator. As a result, Poisson's equation can be easily reformulated as a combination of these constituent neural network, namely, $\mathcal{F}^{\nabla}\left(\mathcal{G}^{\boldsymbol{\kappa}(\{T\})} \cdot \mathcal{H}^{\nabla\{T\}}\right) = \{f(\boldsymbol{x})\}$. It can be easily recognized that the left-hand side is nothing but a larger neural network. Therefore, the Poisson's equation can be represented as a generalized neural network,

$$\mathcal{F}[\{T(\boldsymbol{x})\}] = \{f(\boldsymbol{x})\}$$

Here $\{T(\boldsymbol{x})\}$ and $\{f(\boldsymbol{x})\}$ are lists of physical field values sufficiently sampled from the simulation domain. The above equation is theoretically correct but cannot be used directly, because the size of $\{T(x)\}$ can increase indefinitely as the simulation domain expands, rendering it intractable for training data collecting, neural network construction and training.

To deal with this problem, we can follow the similar procedure of FEM to divide the simulation into subdomains, can assemble the temperature and heat flux values on the nodal points of the subdomains into the vector of $\{T(\boldsymbol{x})\}$ and $\{f(\boldsymbol{x})\}$. The connectivity of the subdomains can be represented as a graph $\mathcal{G}(n, z)$, where $n$ is the number of nodes on the graph and $z$ is the number of links. In fact, treat each element of the FEM mesh as a subdomain, $\mathcal{G}$ can be obtained from the stiffness matrix **K**, where the nonzero slots of **K**form the adjacency matrix of $\mathcal{G}$. Fig. 1(a)-(b) shows an example of triangular mesh and the normalized stiffness matrix (representing nonzero entry by one), while Fig. 1(c) is the corresponding $\mathcal{G}(n, z)$. FEM mesh is an assembly of elements, we can similarly regard $\mathcal{G}$ as the integration of a number of graph communities and each mesh element corresponds uniquely to a graph community. Denoting the node set of an element by $\{\boldsymbol{x}_\alpha, \boldsymbol{x}_\beta, \cdots, \boldsymbol{x}_\eta\}$, the temperature and heat flux on these nodes can be written as

$$\{T\} = \{T_\alpha, T_\beta, \cdots, T_\eta\}, \{q\} = \{q_\alpha, q_\beta, \cdots, q_\eta\}$$

We can easily establish an elementary ANN $\mathcal{A}^e$ to map from $\{T\}$ to $\{q\}$ as illustrated in Fig. 1(d). Consequently, $\mathcal{G}$ can be regarded as the graphical collection of either the input layers or the output layers of elementary ANN models. We assume that all the subdomains are of equal-sized 2D squares or 3D cubes, then the subdomains are of the same phase constituent and microstructure can share an elementary ANN model. Fig. 1(e) shows the schematic of mapping from the temperature graph to the heat flux graph of a uniform material domain via a single elementary ANN. For heterogeneous composite domain but with locally uniform or globally periodic microstructure, only a few elementary ANNs is sufficient to model the physical behavior of entire domain, which drastically reduces the computational and storage costs.

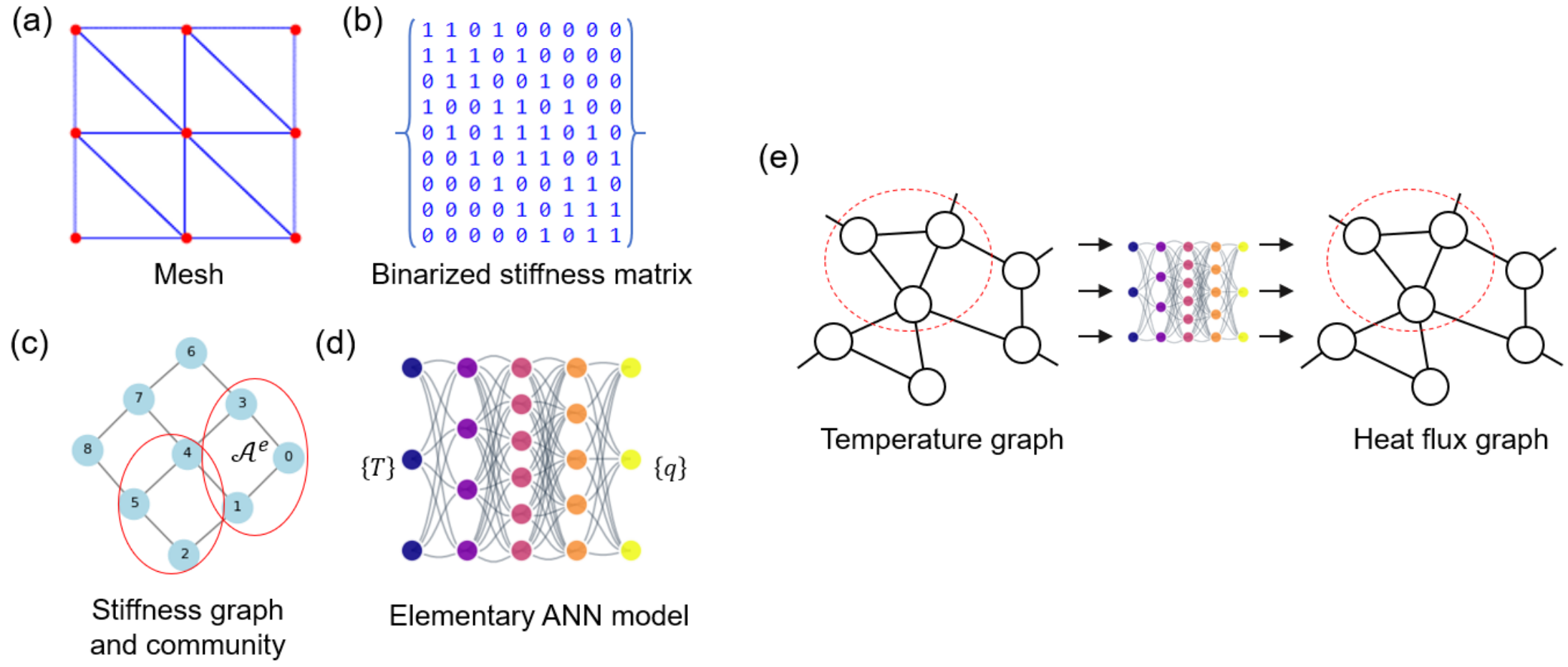

**Figure 1**. (a) a triangular mesh of a small domain. (b) the binarized thermal conduction stiffness matrix of domain (a). (c) the graphical representation of (b) and the illustration of graph communities. (d) the elementary ANN model $\mathcal{A}^e$. (e) the integration of heat graph, elementary ANN, and heat flux graph.

In general, the primary procedure of NEGM is as follows. Let us consider a composite domain with $\chi$ phases which is partitioned $m$ subdomains with a total of $n$ corner points, the number of elements of each phase is denoted by $m_{i\in\chi}$. To simulate the thermal behavior of such composite material via the NEGM framework, we first construct a graph $\mathcal{G}(n, m)$ representing the connectivity of the corner points according to the binarized stiffness matrix. Then, we prepare an input layer of $n$ neurons representing the temperature values of all the partitioned corner points. The input vector will be referred to as $\{T\}_n$. Next, the input vector is divided into small sets according to the community partition of graph $\mathcal{G}$, and these divided sets are batched and stacked into 2D tensors. Afterward, the 2D tensors are fed into the input layers of the corresponding elementary ANNs which produce the 2D tensors of heat flux. Further, we merge the flux tensors according to graph $\mathcal{G}$ to generate the global heat flux vector of the corner points, which is basically the inverse process of temperature dispatch. Finally, boundary conditions (BCs) and physics constraints are enforced to results in a total loss $\mathfrak{L}$. The complete framework of NEGM is demonstrated in Fig. 2. Going through this procedure, we can see that the geometry captured by graph $\mathcal{G}$ and material behavior described by the elementary ANNs are entirely decoupled, which renders NEGM fully scalable with problem size.

The loss term $\mathfrak{L}$ consists of three parts. First of all, the Dirichlet BCs and Neumann BCs are implemented by enforcing the corresponding nodal values of the temperature graph and heat flux graph. Then, energy conservation is respected by zeroing out the summation of the heat fluxes of each graph community. The third part of the loss comes from the governing physics. Specifically, apart from the ANN model mapping from temperature to heat flux (denoted by $\mathcal{K}$), we simultaneous construct and train an auxiliary model (denoted by $\mathcal{K}'$) which maps from heat flux $\{T\}$ to temperature $\{q\}$, so that we can take full advantage of the training data and enhance the convergence of NEGM inference. For Poisson's equation, given temperature distribution $T(\boldsymbol{x})$, the heat flux distribution $q(\boldsymbol{x})$ can be uniquely determined by Fourier's law. Conversely, given the heat flux distribution $q(\boldsymbol{x})$, the temperature $T(\boldsymbol{x})$ is determined up to an additive constant, because the gradient of a scalar field does not fix its absolute value. Therefore, the input of the $\mathcal{K}'$ should include not only the heat flux vector, but a part of the nodal temperature vector to eliminate the additive constant. Specifically, $\mathcal{K}'$ is defined as $\mathcal{K}': \left(\hat{T}\{\boldsymbol{x}\}, q\{\boldsymbol{x}\}\right) \rightarrow T\{\boldsymbol{x}\}$, where $\hat{T}(\boldsymbol{x})$ is a subset of $T(\boldsymbol{x})$. For 2D subdomain, $\hat{T}(\boldsymbol{x})$ can be set to the temperature of a corner point or the temperatures along an edge. Based on the definition of inverse mapping $\mathcal{K}'$, we introduce the following loss term for each element, $\left[\{T\} - \mathcal{K}'\left(\{\hat{T}\}, \mathcal{K}(\{T\})\ \right)\right]^2$. In fact, this loss term provides extra momentum when the backpropagation is stuck in local minimal, so that NEGM inference can more easily achieve the global minimum (see the supplementary material). In sum, the total loss of the NEGM consists of BCs (referred to as "conformation loss", denoted by $\mathfrak{L}_1$), the physical conservation (referred to as "conservation loss", denoted by $\mathfrak{L}_2$), and the physical consistency of reverse mapping (referred to as "consistency loss", denoted by $\mathfrak{L}_3$). In sum, the loss term can be expressed as follows.

$$\mathfrak{L} = \mathfrak{L}_1 + \mathfrak{L}_2 + \mathfrak{L}_3$$

$$\mathfrak{L}_1 = \sum\nolimits_{i \in \Omega(T)} (T_i - \bar{T}_i)^2 + \sum\nolimits_{j \in \Omega(q)} \left(\mathcal{K}(\{T\})_j - \bar{q}_j\right)^2$$

$$\mathfrak{L}_2 = \sum\nolimits_k \left(\sum \mathcal{K}(\{T\}_k)\right)^2$$

$$\mathfrak{L}_3 = \sum\nolimits_k \left[\{T\}_k - \mathcal{K}'\left(\{\hat{T}\}_k, \mathcal{K}(\{T\}_k)\ \right)\right]^2$$

Here, $\Omega(T)$ and $\Omega(q)$ represent the Dirichlet and Neumann BCs, respectively.

The major strength of NEGM is that it completely decouples the mesh and material behavior. First, compared to classical FEM which requires explicit assembly of all the local material behaviors into a global stiffness matrix, NEGM avoids the construction of a global relation by separating the global structure and the local material behavior, leading to a more compact formulation and less storage requirement. Second, compared to the end-to-end ANN model which directly predicts a global field, NEGM is not only fully scalable but also training friendly since the only trainable parts are the small sized elementary ANNs and their inverse counterparts. Once the elementary ANNs are trained properly, the parameters of elementary ANNs are fixed while the global input becomes unknown. Given a meshed composite domain under given BCs, its solution can be obtained by inference of the corresponding NEGM, which is equivalent to train for the input layer. Compared to the popular combination of domain decomposition and Schwarz iteration[32], NEGM is basically a giant ANN but with fixed hidden parameters, making it much more efficient owing to the gradient back-propagation technique, as implied in Fig. 2. The major difference between classical numerical method, such as FEM, is that solving by NEGM is purely based on the gradient backpropagation of a couple of small ANNs instead of a giant global stiffness constructed by numerical differentiation. This make NEGM both light-weight and computationally efficient.

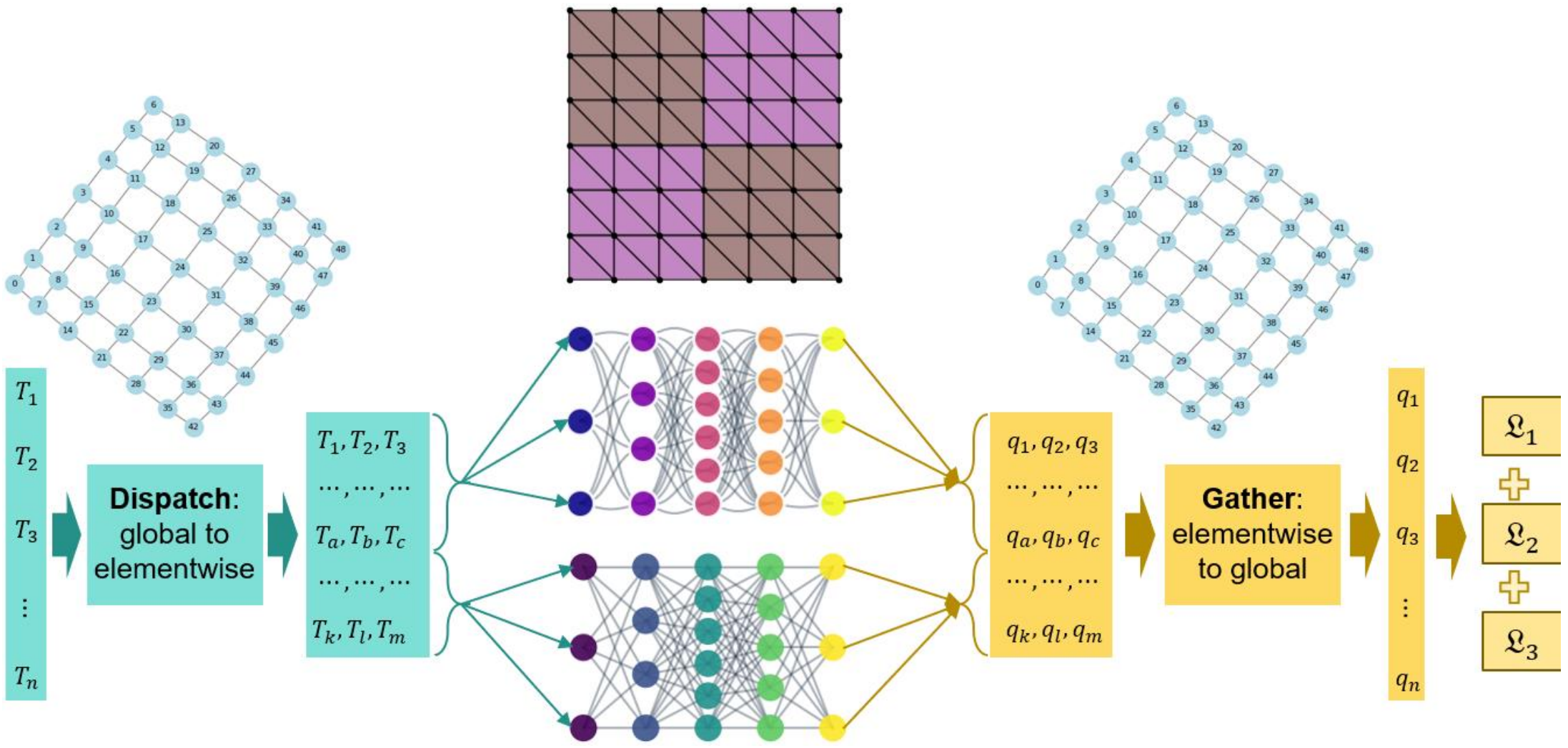

**Figure 2**. The work flow of NEGM. First, it dispatches the global temperature field to the input layer of the elementary ANNs according to the temperature graph; then, it processes the input temperature field by the ANNs and produce elementwise heat flux

tensor; next, it gathers the global heat flux field according to the heat flux graph; finally, it calculates the loss function by enforcing BCs and physical consistency. The calculated loss is back propagated to update the global temperature field until convergence.

### 3. Validations on nonlinear material in 2D

To validate the primary idea of NEGM, we first apply it to the thermal conduction simulation of uniform nonlinear 2D material and compare its inference results with FEM results. As shown in Fig. 3(a), four different types of nonlinear material models are synthesized with strongly different dependences of thermal conductivity on temperature. For each material, we construct a representative quadrilateral element of unit length along each dimension and build a pair of multi-layer perceptron (MLP) (a forward one and a backward one) to describe its thermal behavior, i.e., mapping from temperature to heat flux and vice versa, as shown in Fig. 3(b). The MLPs of each material is trained by 5000 sets of randomly generated nodal temperatures and the corresponding heat fluxes calculated by FEM. For each material type, we prepare a simulation problem over a slab of $10 \times 10$ and partition it into unit-size cells, as shown in Fig. 3(c). We fix the temperature of the right boundary of the domain and apply a uniformly distributed heat source on the left boundary. For each simulation case, we first employ NEGM to calculate the temperature field, and then use FEM to obtain the ground truth by meshing the domain into $50 \times 50$ quadrilateral elements.

The simulation results of steady-state temperature by NEGM and FEM are demonstrated in Fig. 3(d). To quantitatively evaluate their discrepancies, we have computed the average temperature profile along the direction of applied heat flux for each simulation case. As can be seen, the NEGM-calculated temperature profiles are in perfect agreement with the FEM results, with an averaged relative error less than 1% for all the simulation cases. These results suggest that the NEGM framework is algorithmically correct in dealing with the physical simulation nonlinear materials by decoupling the complexity in material behavior and the geometrical partition of simulation domain. In practical applications, the proposed NEGM algorithm can also deal with complex microstructures and exotic simulation domain geometries in 3D space under arbitrarily set BCs, as we will see in the following sections.

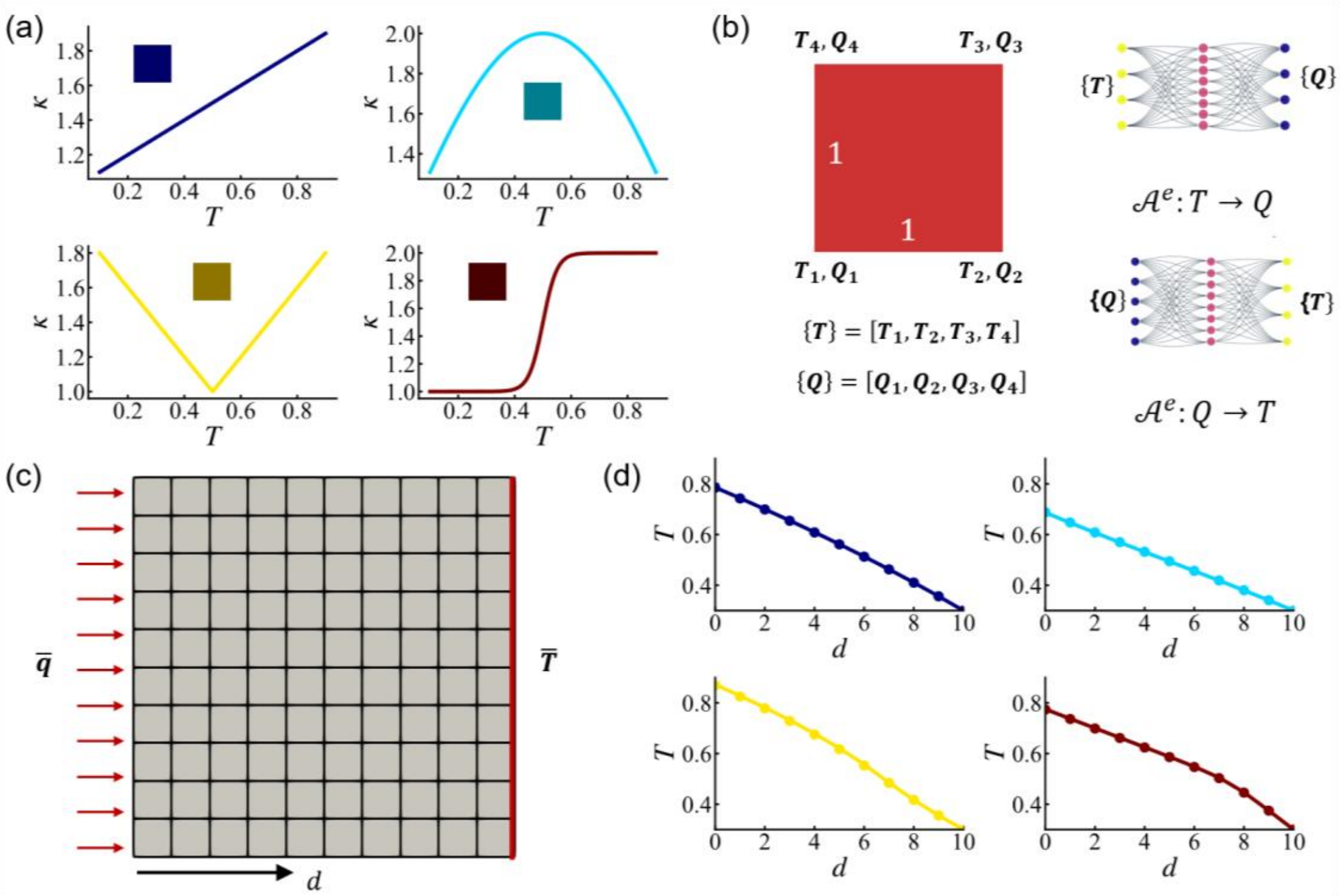

**Figure 3**. Validation of NEGM on 2D simulation domains of nonlinear materials. (a) the thermal conductivity vs. temperature relations of the four different materials. (b) definitions of the unit cell and the forward and backward MLPs. (c) construction of the simulation domain and the boundary conditions, where $\bar{T} = 0.3$ and $\bar{q} = 0.15$. (d) comparison of the simulation results obtained by NEGM and FEM. For FEM simulation, each unit cell of the simulation domain is meshed into $5 \times 5$ quadrilateral elements. The comparisons are done for the average temperature profiles along the heat

flux direction. The scattered the points are results from NEGM while the lines are from FEM.

As discussed in previous sections, the total loss function of the NEGM consists of three components, the consistency error, the conservation error, and the BC error. During the inference process, these three components need to evolve jointly to achieve global convergence. We have also prepared an example to investigate how they evolve during the running of NEGM. As shown in Fig. 4(a)-(b), the domain is still $10 \times 10$ and purely made of the second type of material. The applied BCs is illustrated in Fig. 4(c), where the heat fluxes with magnitude of $0.05$ are injected through the central areas of the left and top boundaries, and fixed temperatures of $0.3$ are enforced up the central areas of the right and bottom boundaries. When employ NEGM, the temperature of the entire simulation domain is initialized to $0.4$. Fig. 4(d) demonstrates the temperature field evolution with respect to the iteration steps of NEGM. As we can see, the temperature field changes dramatically in the initial inference stage. Afterward, it evolves gradually to reach the final high-precision solution.

Fig. 4(e) demonstrates the evolutions of these components with respect to the iteration steps. As we can see, the three components behave rather differently during the inference process. The BC error monotonically decreases from a large initial state to reach a sufficiently small state; the conservation error initially drops sharply, subsequently rises to a higher level, and then decreases steadily from there; the consistency error maintains a small level during the whole inference process. These results suggest that the dominant objective of NEGM is the reduction of BC error, while the conservation and consistency errors are of secondary importance. The conservation and consistency loss terms primarily contribute to maintaining the physical consistency of the inferred solutions and accelerating convergence during training. Consequently, they act as regularization mechanisms that smooth the optimization landscape, thereby promoting a flatter loss surface and more stable convergence behavior.

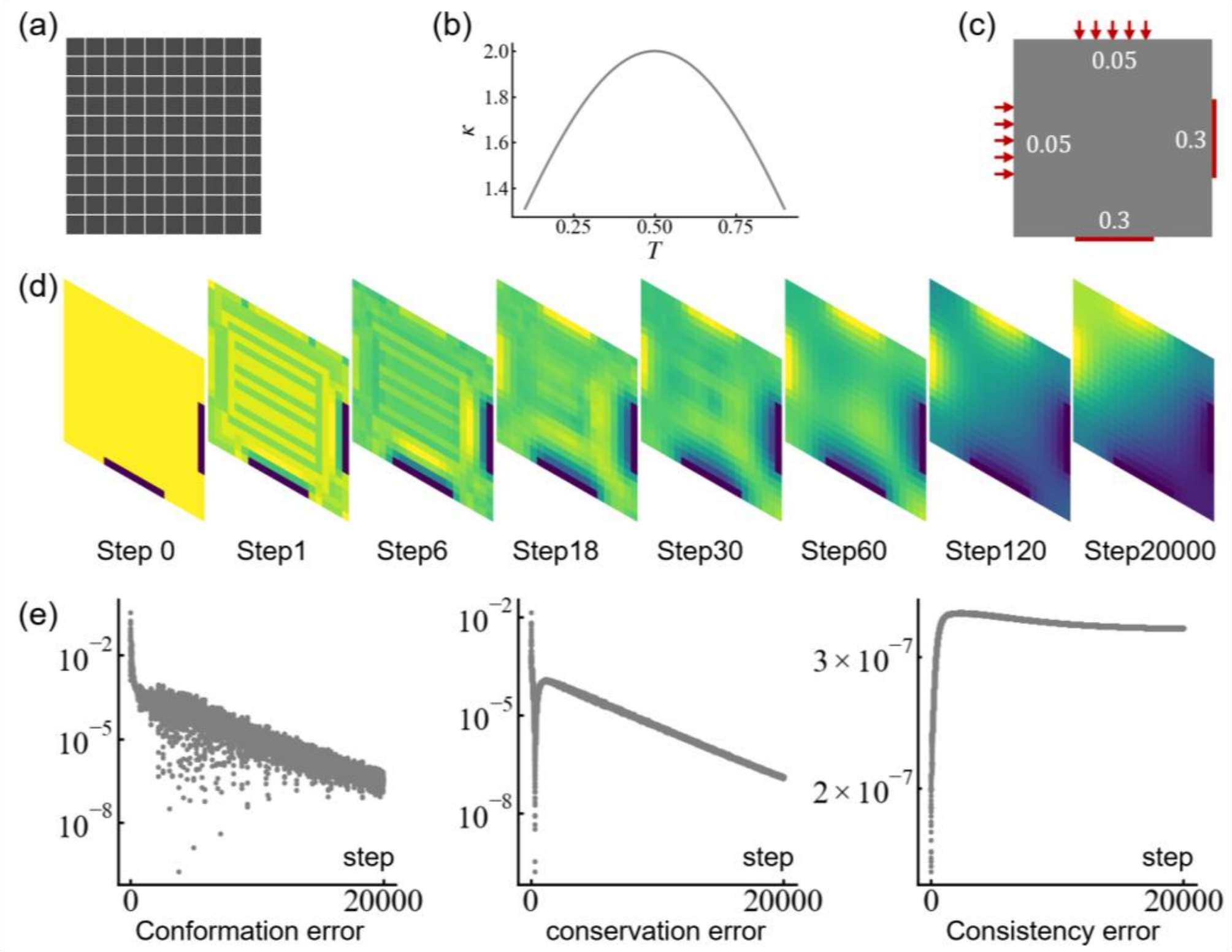

**Figure 4**. (a) the simulation domain to investigate the evolution of loss components during NEGM inference. (b) the nonlinear thermal conductivity to be assigned to the domain. (c) the applied temperature and heat flux boundary conditions, where the central areas of left and top sides are prescribed with heat flux of $+0.05$ and the central areas of right and bottom sides are prescribed with temperature of $0.3$. (d) the evolution of the temperature field within the domain during the NEGM inference. (e) the evolutions of conformation loss, conservation loss, and consistency loss, during the inference process.

## 4. Simulation of heterogeneous composite in 3D

Although ANN-based physical modeling has been intensively investigated in recent years, many such models remain constrained to 2D problems. For 3D, the curse of dimensionality presents an unavoidable obstacle that usually causes computational cost to explode. Here, we will demonstrate that the proposed NEGM can easily deal with the simulation of 3D composite material. Instead of constructing 2D quadrilateral representative element, now we directly consider 3D cubic cells for each material, as illustrated in Fig. 5(a)-(b). The material models of the cubic cells are identical with those in previous 2D simulation cases. Based on these cubic building cells, we can construct complex 3D microstructure and irregular 3D simulation domains. For each elementary cell, the corner points are indexed and listed as 8-element vector to feed the forward and backward MLP mappings, as illustrated in Fig. 5(c)-(d). Training of the MLPs is similar as in previous section. We randomly assign temperature values to the corner points of the cubic cell, and use FEM to calculate the corresponding heat powers on each corner nodes. 10000 sets of such data entries are prepared for each material cell.

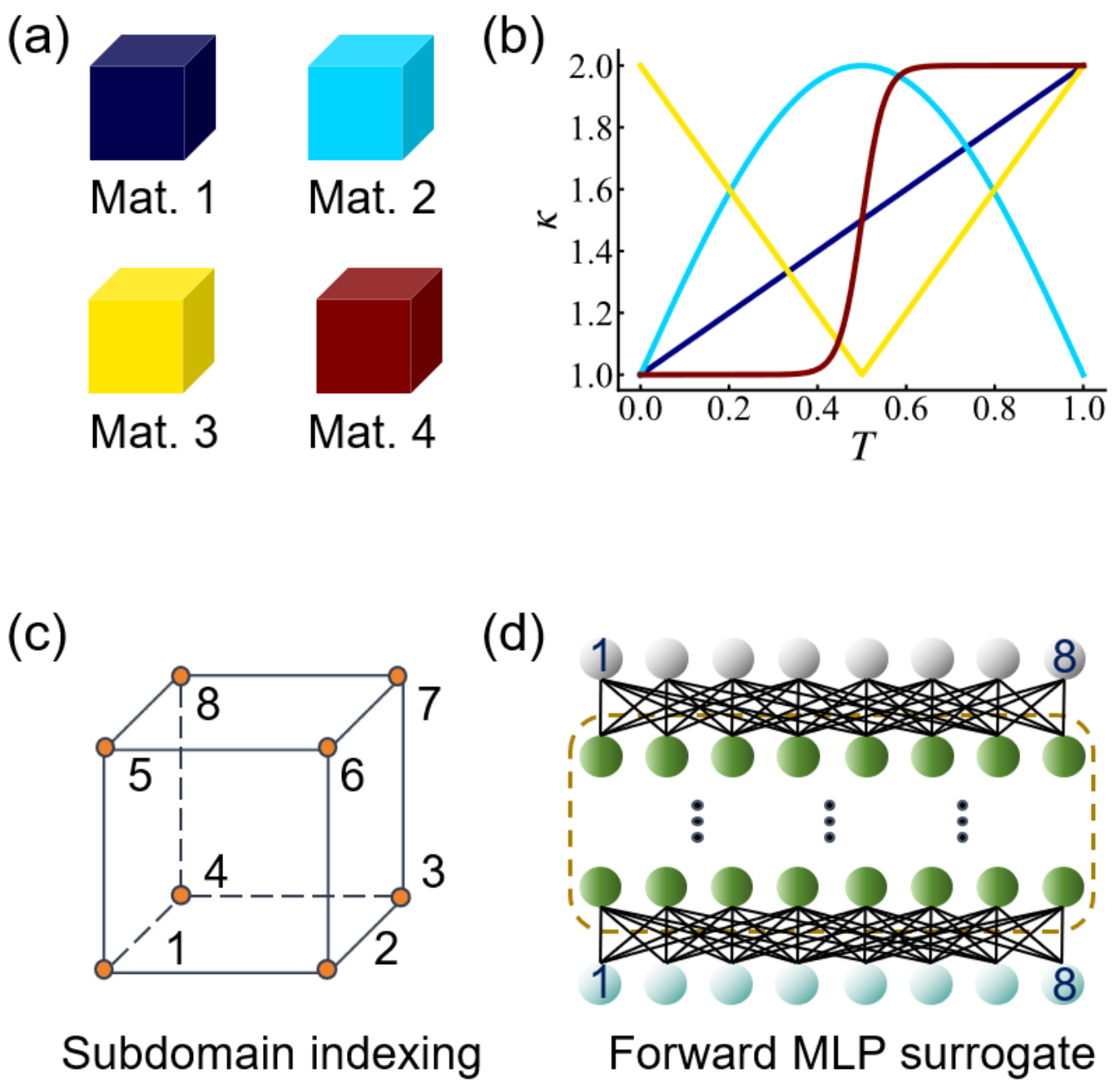

**Figure 5**. (a)-(b) the cubic building-block cells and thermal conductivity models of the four materials used to construct 3D composite domain. (c)-(d) the indexing of corner nodes and the general architecture of the MLP model mapping from nodal temperature to nodal heat power. Similarly, the input and output dimension of the backward MLP is 9 and 8, respectively.

Our first example is a synthesized composite consisting of the four material cells that are randomly distributed, as demonstrated in Fig. 6(a). Although machine-learning-based solution of random composite have be extensively explored, lots of them are constrained to simple BCs, such as pure Dirichlet BCs. Here, we employ NEDG to solve a complicated problem involving a mixture of locally concentrated Dirichlet and Neumann BCs. Fig. 6(b) shows the pattern of the applied BCs, where the outline box represents the simulation domain, the surface patches $S_1$, $S_3$, $S_5$ represent the locations where uniform Neumann BCs are applied, while the scatter point groups $S_2$, $S_4$, $S_6$ represent the locations where node-wise Dirichlet BCs are enforced. As can be seen, this simulation case simultaneously involves the coupled complexities of heterogeneous microstructure and mixed BC types.

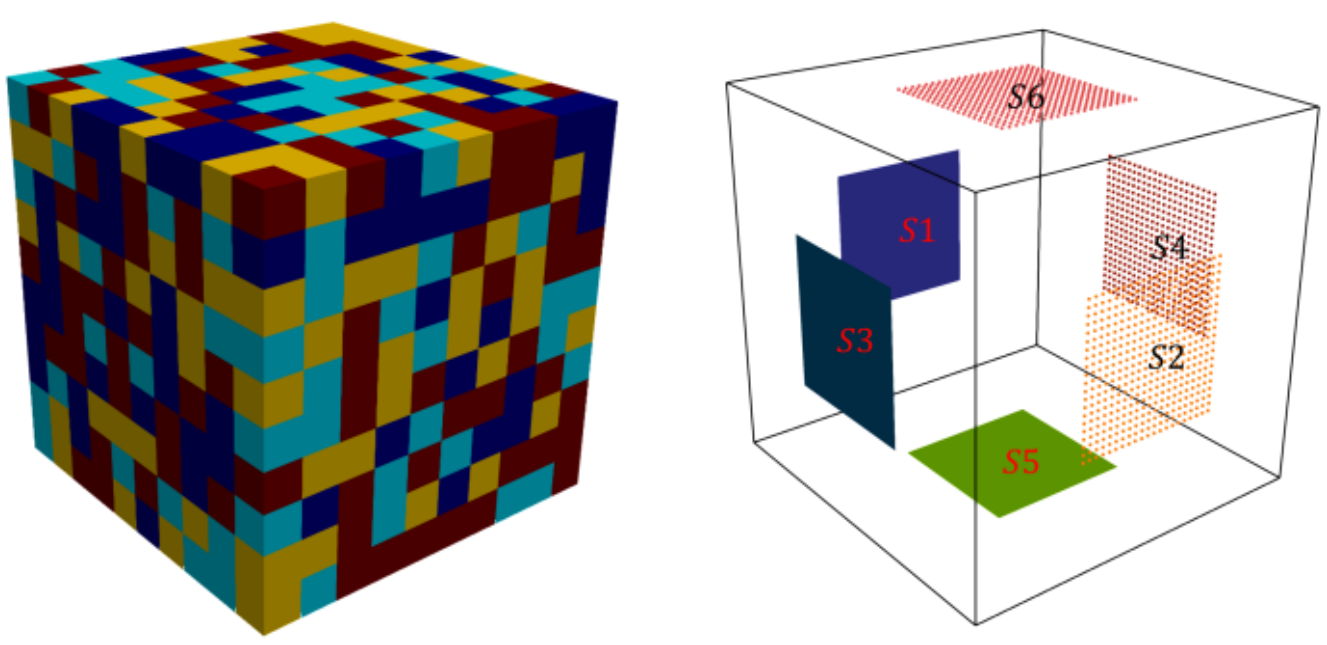

**Figure 6**. The microstructure configuration of the 3D composite domain and the mixed boundary conditions. For Neumann BCs, $S_1$, $S_3$, and $S_5$ specify the heat flux with magnitude of $+0.15$; For Dirichlet BCs, $S_2$, $S_4$, and $S_6$ specify the temperature with magnitude of 0.3.

To evaluate the performance of NEGM in solving this case, FEM calculation for this problem is also carried out by finely meshing the domain into $50 \times 50 \times 50$ brick elements and the FEM result is regarded as ground truth. Fig. 7(a)-(b) compares the temperature fields obtained by NEGM and FEM. As we can see, the two results are of nearly identical pattern, suggesting that NEGM have successfully captured the physical process of thermal conduction. Quantitatively, according to the error map shown in Fig. 7(c), the largest relative discrepancy between the NEGM prediction and the FEM ground truth is less than $6\%$. Fig. 7(d) is the evolution of loss function $\mathfrak{L}$ during NEGM inference iterations, which shows that the total loss converges at about $5000$ iterations. The salient convergence point suggests that the convergence of NEGM inference is well-defined. Fig. 7(e) demonstrates a point-by-point comparison between the NEGM result and the FEM result at the cell corner points, where the average discrepancy is $2\%$, and the standard deviation is $0.6\%$. This validation study confirms the versatility of NEGM as a general-purpose solver for both two-dimensional and three-dimensional composite simulations. The results indicate that NEGM can achieve computational efficiency and physical fidelity comparable to those of classical FEM solvers.

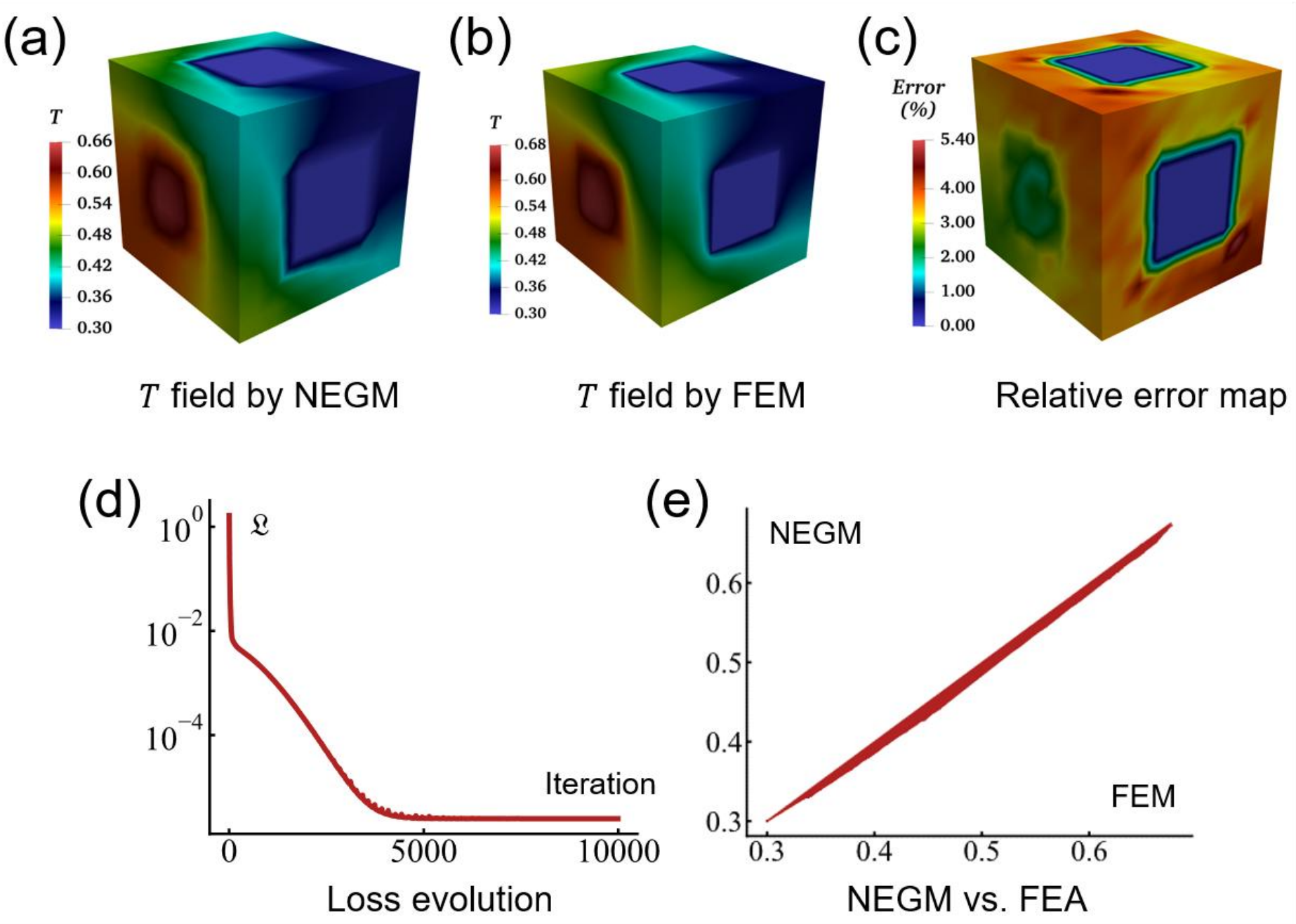

**Figure 7**. (a) the NEGM simulation result over the $10 \times 10 \times 10$ random composite domain. (b) the FEM result of the random composite mesh into $50 \times 50 \times 50$ brick elements. (c) the relative error map of the NEGM result with respect to the FEM result. (d) the evolution of loss $\mathfrak{L}$ against NEGM inference iteration. (e) the point-by-point correspondence between NEGM and FEM results.

Apart from dealing with complex microstructure and mixed BCs, NEGM can also deal with irregular simulation domain. Fig. 8(a) shows an exotic shaped domain is similar as a heat radiator consisting of a base and a couple of pins. It is made of the four nonlinear materials shown in Fig. 5(b). A constant heat flux of $+0.02$ is uniformly enforced on the bottom face of the base and a fixed temperature of $0.3$ is enforced on all the top faces of all the pins. Fig. 8(b)-(c) shows the cell configuration for NEGM and mesh used by FEM, while Fig. 8(d)-(e) compares the calculated temperature fields by these two methods. As can be seen, NEGM has accurately captured the temperature distribution pattern and the relative discrepancy of the largest temperature in NEGM result is approximately $0.3\%$. This example demonstrates that the proposed NEGM can reliably accommodate irregular shaped simulation domains, thereby exhibiting an important characteristic expected of a general-purpose solver. It can be employed to deal with the physical modeling of arbitrary structural component, just like classical FEM.

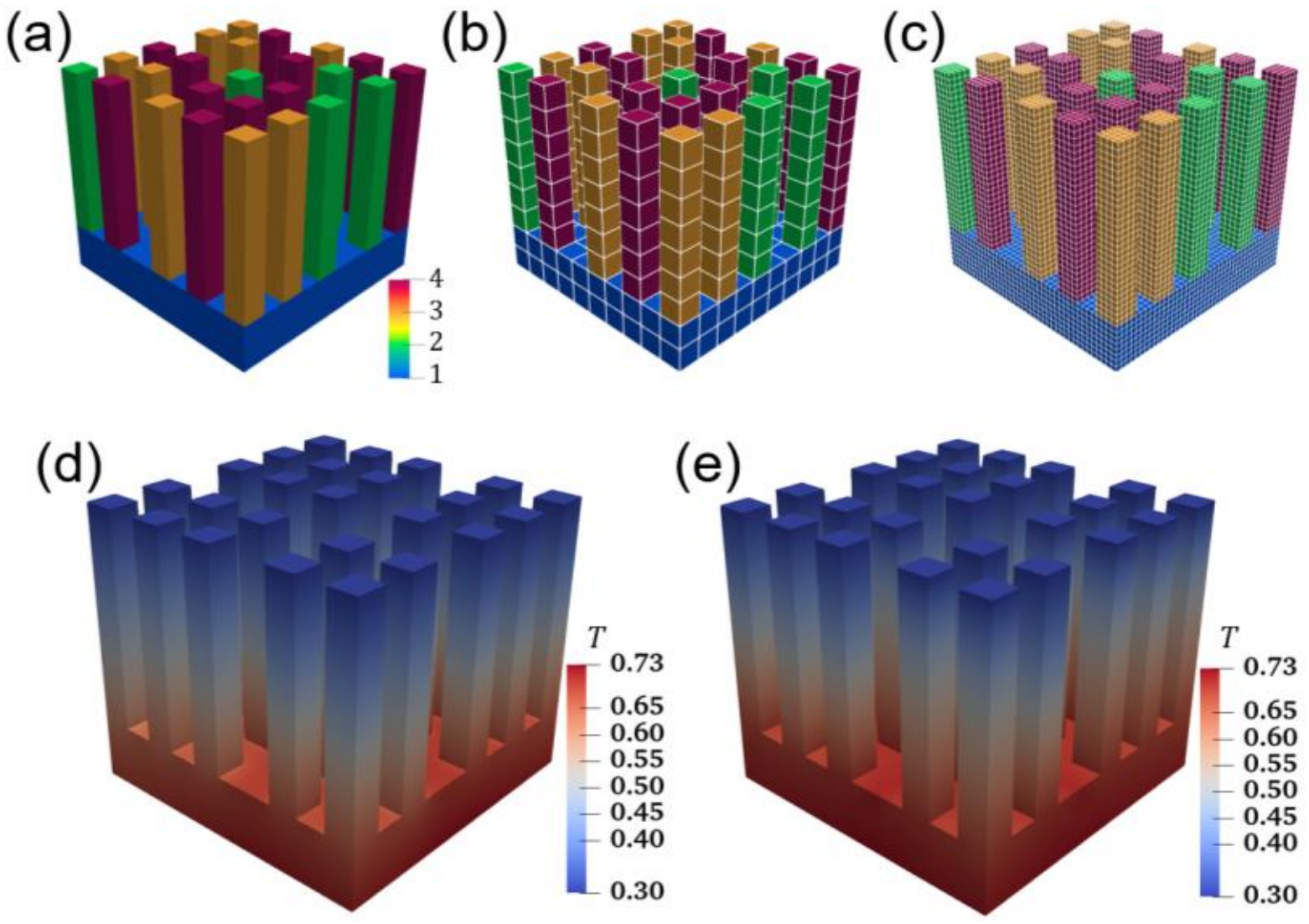

**Figure 8**. A test case for NEGM to simulate the thermal conduction problem of a heat-radiator shaped composite domain of nonlinear materials in 3D. (a) the domain geometry and the material distribution within. It consists of 4 different materials, whose thermal conductivities are shown in Fig. 5(b). A uniform heat flux with magnitude of 0.02 is injected through the bottom face of the domain, while the top faces of the pins are prescribed by fixed temperature of 0.3. (b) the cell configuration for NEGM. (c) the mesh used by FEM. (d) the FEM simulation result of temperature field. (e) the NEGM prediction result of temperature.

## 5. Self-consistent hierarchical upscaling

Previous sections have proven that the proposed NEGM framework can easily handle the simulation involving strongly nonlinear materials, arbitrarily mixed BCs, exotic domain geometries in 2D and 3D. However, as the size of the simulation domain increases, the computational time and memory requirements of NEGM also grow substantially, eventually becoming prohibitively expensive for large-scale problems. In conventional FEM, this challenge is typically addressed by leveraging high-performance computing resources and distributed parallel computing. Although such approaches can alleviate the computational burden, they fundamentally rely on increasing hardware resources, resulting in higher computational costs, while the achievable parallel efficiency is often constrained by communication overhead and load imbalance. NEGM provides another pathway to solve such situation. Since the primary constituents of NEGM are the small building-block ANNs and NEGM itself is a large neural network, it provides a unified framework to deal with the simulation at arbitrary length scale. Although the building-block MLPs in above examples are trained by FEM results, it is equally legitimate to train them with the inference results of NEGM. Therefore, we can take a progressive upscaling strategy by collecting data via NEGM inferencing and training higher level MLPs with NEGM data.

To elaborate this idea, let us consider the 2D mosaic composite made of the four nonlinear materials. The mosaic composite contains $1000 \times 1000$ alternatingly patched cells of the four materials as described. Our goal is to determine the steady-state temperature over this simulation domain under a given set of BCs. Although NEGM can, in principle, be applied directly to solve this problem, the presence of one million cells still poses a significant challenge to its convergence. Instead, we can take the following progressive upscaling procedure. First, we construct a $10 \times 10$ mosaic composite domain $\mathcal{D}_{10}$ which is of the same cell arrangement as the original $1000 \times 1000$ composite domain. For this small system, NEGM can easily calculate its response to any BCS, which enables us to efficiently collect large amount of behavioral data under various BCs and learn a pair of MLPs (denoted by $\mathcal{F}_{10}$) for the entire domain $\mathcal{D}_{10}$. Next, we deal with a $100 \times 100$ mosaic composite domain $\mathcal{D}_{100}$ by treating it a $10 \times 10$ domain of the $\mathcal{D}_{100}$ cells and construct a NEGM based on the $\mathcal{F}_{10}$. We use this NEGM to collect the behavioral data under randomly specified BCs and learn a pair of MLP (denoted by $\mathcal{F}_{100}$) for the entire domain $\mathcal{D}_{100}$. Finally, we can address the original simulation problem of $1000 \times 1000$ composite domain by treating it as $10 \times 10$ domain of the $\mathcal{D}_{100}$ cells. We construct a NEGM based on the $\mathcal{F}_{100}$ to determine the thermal response of the $1000 \times 1000$ composite under given BCs. For comparison, we have also performed the direct FEM calculation for the $1000 \times 1000$ composite domain, whose result is served as the ground truth. Fig. 9 demonstrates the work flow of the progressive upscaling. As can be seen, the temperature fields obtained by progressive NEGM upscaling and the direct FEM are in perfect agreement.

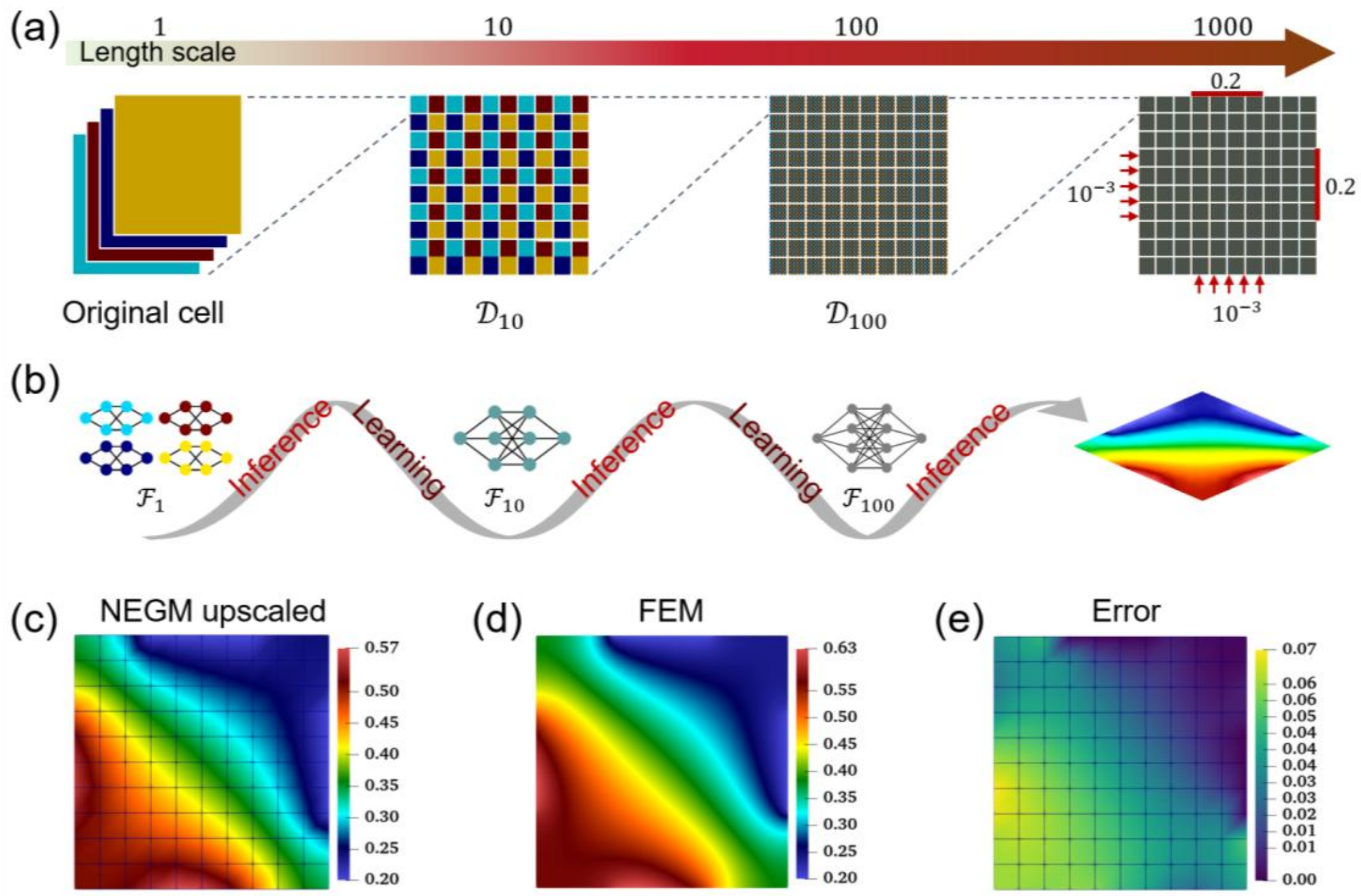

**Figure 9**. Schematic illustration of progressive NEGM upscaling. (a) divide the simulation for a large composite domain into three steps at length scale 10, 100, and 1000. $\mathcal{D}_{10}$ and $\mathcal{D}_{100}$ denote the representative domains at the corresponding length

scales. The target is to solve a $1000 \times 1000$ domain under prescribed temperature of 0.2 at the central areas of the top and right edges and uniform heat flux of $10^{-3}$ at the central areas of bottom and left edges. (b) train MLPs for the representative domains at length scale $1$, $10$, and $100$. (c)-(d) the temperature fields obtained by NEGM upscaling and direct FEM. (e) the discrepancy of NEGM upscaling result compared to FEM ground truth.

It can be recognized that the above upscaling method is actually a statistical coarsening process, which is reminiscent of the renormalization group (RG) flow in statistical physics. The RG flow is dominated by the coarsening and renormalization operations. For a given material domain at length scale $\theta$, its physical response to specific BCs $(\overline{\boldsymbol{p}}, \overline{\boldsymbol{q}})$ is an invariant. To obtain the response using NEGM, we need to determine the representative cell and train the MLPs for it. For a representative cell at length scale $\zeta$, we denote the MLPs by $\mathcal{F}_\zeta$ and the NEGM by $\mathfrak{G}^\theta(\mathcal{F}_\zeta)$; if the representative cell is of length scale $\eta$ ($\theta > \eta > \zeta$), then the MLPs and the NEGM can be denoted by $\mathcal{F}_\eta$ and $\mathfrak{G}^\theta(\mathcal{F}_\eta)$. Denoting the MLP learning by $\gg$, the coarsening and renormalization operations can be described as $\mathfrak{G}^\theta(\mathcal{F}_\zeta, \overline{\boldsymbol{p}}, \overline{\boldsymbol{q}}) = \mathfrak{G}^\theta(\mathcal{F}_\eta, \overline{\boldsymbol{p}}, \overline{\boldsymbol{q}})$ and $\mathfrak{G}^\eta(\mathcal{F}_\zeta) \overset{\gg}{\rightarrow} \mathcal{F}_\eta$, respectively. The former suggests that the original domain at scale $\theta$ can be equivalently characterized by $\mathfrak{G}^\theta(\mathcal{F}_\zeta)$ or $\mathfrak{G}^\theta(\mathcal{F}_\eta)$, while the latter means that $\mathcal{F}_\eta$ can be obtained by learning from the inference results of $\mathfrak{G}^\eta(\mathcal{F}_\zeta)$, as illustrated in Fig. 10(a)-(b). By iterative employment of coarsening and renormalization, the computational cost of the large-scale modeling can be reduced.

Fig. 10(c)-(d) shows the evolutions of the computational accuracy and efficiency during the upscaling process. Here, the accuracy is defined as the system-wise averaged relative differences between the upscaled results and the direct FEM results, while the efficiency is defined as the relative time cost of the upscaling method with respect to direct FEM. Note that we do not include the training time for the upscaling method since it is just one-time cost. The accuracy and efficiency are measured over 100 cases under random BCs. As can be seen, the relative error of the self-taught NEGM upscaling method is about 3% after 3 upscaling steps with a clear degradation trend. Such degradation is mainly caused by the accumulation of learning and inference errors along the upscaling steps, which needs to be further investigated. In terms of efficiency, at the initial step, the cost of NEGM is three times that of FEM. As upscaling goes on, the former gradually decreases and become negligible just after 3 steps. Since each upscaling step is algorithmically identical, the time cost of the NEGM upscaling scales linearly with the number of steps, which is a great advantage when dealing with for extremely large computational domain.

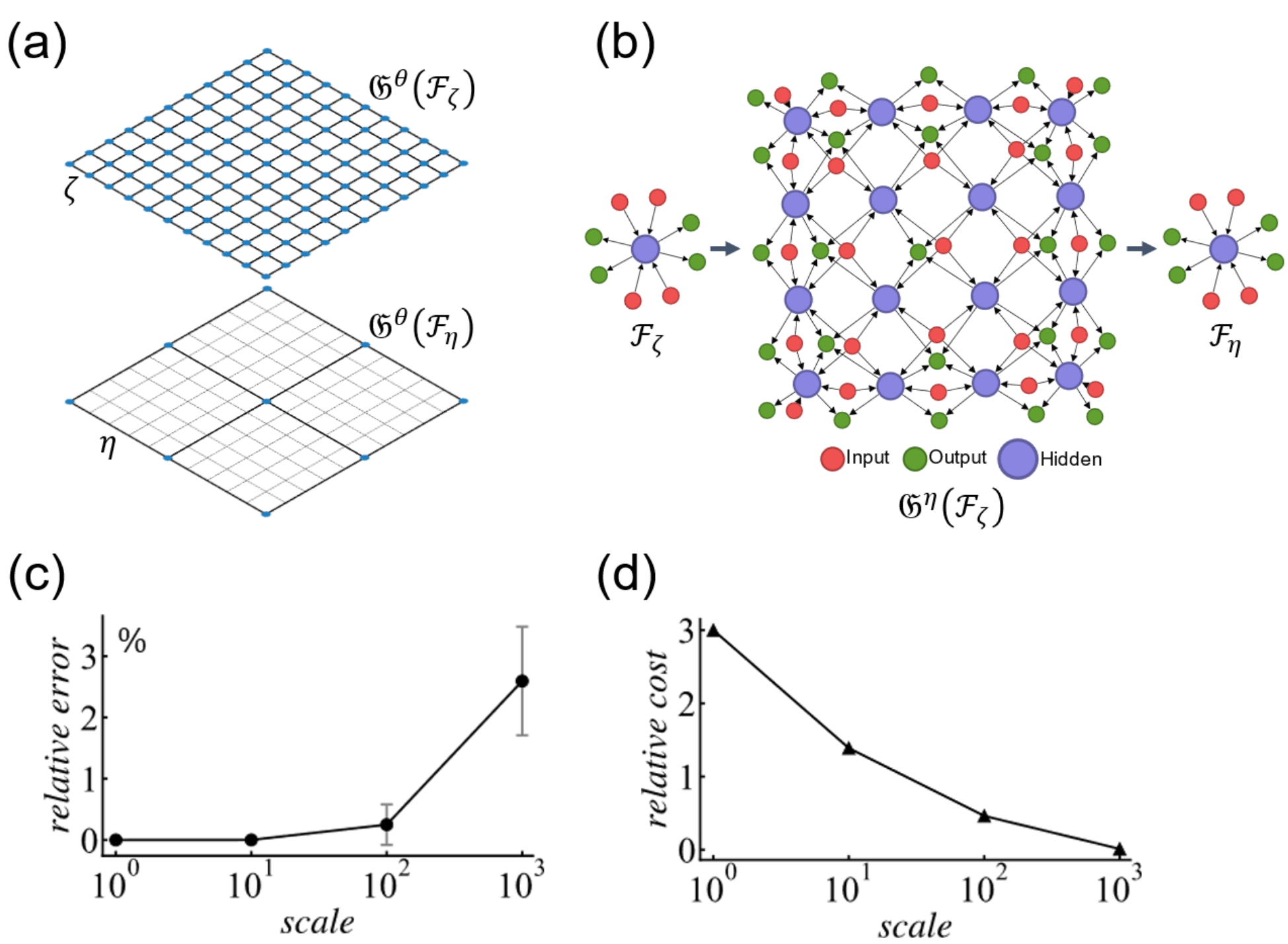

**Figure 10**. Illustration of the correspondence between NEGM upscaling and RG flow. (a) illustration of the coarsening step (NEGM inference), where $\theta$ is the length scale of the original simulation domain, $\zeta$ and $\eta$ are the length scales of the intermedia representative cells with $\eta > \zeta$. (b) illustration of the renormalization step, which trains the MLPs of length scale $\eta$ using the NEGM of length scale $\zeta$. (c) the relative error evolution with respect to NEGM upscaling step averaged over 100 cases. (d) the relative time cost evolution with respect to upscaling step.

As we know, the primary goal of RG in statistical physics is to understand how the behavior of a many-body system changes with the observation scale. More specifically, RG provides a systematic framework for eliminating microscopic degrees of freedom while preserving the macroscopic physics. In our framework, the objective of NEGM upscaling is to progressively coarse-grain the small-scale effects while preserving only the essential features governing the behavior of the original large-scale simulation domain. To better appreciate the coarsening effect of the NEGM upscaling method, Fig. 11(a) demonstrates the density distributions of the trainable weights of the MLPs obtained at different upscaling steps. At small scales, the structural heterogeneity dominates, which requires the MLP contain large number of nonzero coefficients to resolve the structural complexity. As the length scale further increases, the statistical averaging effect comes into play, leading to statistically smoothened thermal behavior and more negligible coefficients. Fig. 11(b) summarizes the number density of negligible coefficients of the MLPs obtained at different length scales, which clearly demonstrates the increasing and saturating trend. Therefore, similar as the RG flow in statistical physics, the self-taught NEGM upscaling method can also suppress the irrelevant parameters and zero out more coefficients in the MLPs as the upscaling proceeds. In general, this upscaling method offers a potential pathway to universal multiscale modeling by repeated coarsening and renormalization, as illustrated in Fig. 11(c). Specifically, given the microstructure of a material (either aerogel, composite, granular, or polycrystals, etc.), we can identify the typical structures and collect their behavioral data by classical computational methods, such as MD or FEM. The collected data is then used to the train the building-block MLP models for the representative cells of these structures. Based on these MLPs, the NEGM upscaling can be repetitively called, until reaching the desired length scale.

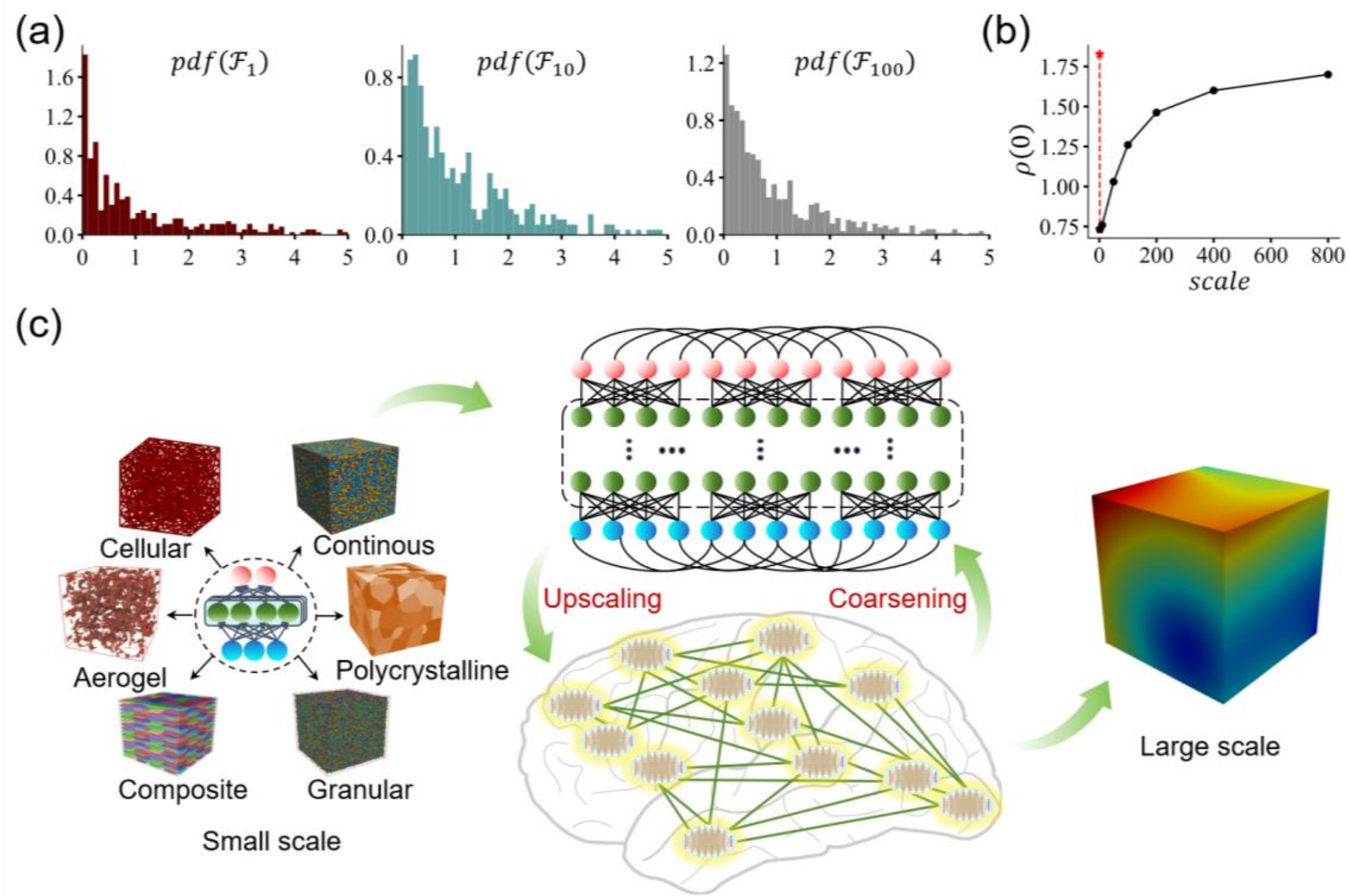

**Figure 11**. The NEGM-based progressive multiscale coarsening. (a) the density distributions of the absolute values of the well-trained MLP coefficients at length scale 1, 10, and 100. (b) the number density of zero coefficients $\rho(0)$ vs. length scale. (c) illustration of deriving the macroscopic behavior from the microscopic structures through the self-taught NEGM upscaling method.

## 6. Conclusion

In this work, we proposed a neural-embedded graphical model (NEGM), a scalable computational framework that addresses one of the fundamental limitations of conventional ANN-based physical modeling: the inability of fixed-dimensional neural networks to naturally accommodate problems with arbitrary spatial scales and geometries. By decoupling geometric representation from constitutive behavior, the proposed framework separates a simulation task into an extensive component, represented by a complex graph describing the domain topology, and an intensive component, represented by reusable building-block neural networks that characterize the local physical response. This decomposition enables the scalability of the overall simulation to be governed by the graph structure rather than the neural network architecture itself.

The proposed framework demonstrates remarkable flexibility in handling arbitrary BCs, irregular 2D and 3D geometries, and simulation domains of variable sizes without requiring retraining of the underlying neural models. Compared with conventional domain decomposition methods, NEGM formulates the entire problem as a unified differentiable graphical model, allowing the global solution to be obtained directly through gradient-based optimization without introducing Schwarz-type iterative coupling between subdomains. Consequently, the inference procedure remains algorithmically simple while preserving broad applicability. More importantly, this work establishes a unified framework for progressive multiscale upscaling. Rather than coupling heterogeneous simulation methodologies across different length scales, NEGM employs self-consistent inference to generate behavioral data that are subsequently used to train reusable building-block neural networks for progressively larger representative cells. This recursive coarse-graining strategy enables large-scale physical behavior to emerge automatically from the responses of smaller-scale constituents, providing a computationally efficient and conceptually unified alternative to conventional multiscale modeling approaches. In this sense, the proposed framework shares the same philosophy as renormalization-group methods by progressively eliminating fine-scale degrees of freedom while preserving the essential macroscopic behavior.

Although the present study focuses on regular square and cubic building-block cells with corner-based physical variables, the framework is readily extensible. Future work will incorporate Jacobian-based coordinate transformations to accommodate arbitrarily shaped cells and develop more expressive building-block neural networks capable of representing heterogeneous multiphase microstructures using enriched sampling strategies. These developments are expected to further enhance the geometric flexibility and physical representability of NEGM, paving the way toward general-purpose, scalable neural simulation of complex engineering materials and structures.

**Data availability**

The data and codes involved in this work are available upon reasonable request.

**Acknowledgement**

This work is supported by the National Natural Science Foundation of China (No. U24A20168, 12302178, 12402162), the Natural Science Foundation of Jiangsu Province (No. BK20230897), and China Post-doctoral Research Foundation (Grant No. 2025M771863).